# Driving conditions dependence of magneto-electroluminescence in tri-(8-hydroxyquinoline)-aluminum based organic light emitting diodes

Qiming Peng, Jixiang Sun, Xianjie Li, Mingliang Li, and Feng Li[a]

State Key Lab of Supramolecular Structure and Materials, Jilin University, 2699 Qianjin Avenue, Changchun 130012, People's Republic of China

**Abstract**: we investigated the magneto-electroluminescence (MEL) in tri-(8-hydroxyquinoline)-aluminum based organic light-emitting diodes (OLEDs) through the steady-state and transient method simultaneously. The MELs show the great different behaviors when we turn the driving condition from a constant voltage to a pulse voltage. For devices driven by the constant voltage, the MELs are similar with the literature data; for devices driven by the pulse voltage, the MELs are quite different, they firstly increase to a maximum then decrease as the magnetic field increases continuously. Negative MELs can be seen when both the magnetic field and driving voltage are high enough.

[a] Author to whom correspondence should be addressed; Electronic mail: lifeng01@jlu.edu.cn

The magnetic field effects (MFEs) on electroluminescence (namely MEL) and current (namely MC) in organic semiconductor devices with non-magnetic electrodes have attracted much attention in recent years.[1-25] Kalinowski *et al.* discovered that the electroluminescence (EL) and current of tri-(8-hydroxyquinoline)-aluminum based organic light-emitting diodes (OLEDs) can be enhanced by a modest external magnetic field.[1] They attributed it to the magnetic-field-modified intersystem conversion (ISC) between the singlet and triplet electron-hole pairs (e-h pairs). Mermer *et al.* investigated a number of organic semiconductor devices and observed both positive and negative MCs.[2] In addition, they reported that the MFEs can be observed even in hole-only-transporting device.[3] Large MEL (> 50%) was observed by Nguyen *et al.*.[4] These authors suggested that the MFEs should be explained by the bipolaron model,[4,5] which was proposed by Bobbert *et al.*.[6] Meanwhile, Desai *et al.* investigated the MFEs on the current transporting in the devices under illumination and suggested the triplet-polaron interaction (TPI) model should act as a major mechanism to explain the MFEs.[7,8] Davis *et al.* and Xiong *et al.* suggested that the influence of the magnetic field on the triplet-triplet annihilation (TTA) should account for the MFEs.[9-11] Hu *et al.* proposed that the primary carriers (carriers injected from the electrodes) and the secondary carriers (carriers resulted from the dissociation of excitons and/or from the exciton-carrier interaction) play different roles in the MFEs.[15] More recently, Niedermeier *et al.* reported an enhancement of the MFEs when the devices were electrically stressed for a few minutes before taking the MFEs measurement.[12,13]

It has to be mentioned that for the similar devices tested under the similar driving conditions, various valves of MEL (from <3% to >30%) and MC (from <1% to > 15%) were reported.[4,13,14] To identify the microscopic mechanisms of MFEs in organic semiconductors, most studies focused on the steady-state method (OLEDs driven by the constant voltages or currents),[1-20] only a few works used the transient method (OLEDs driven by the pulse voltages).[21-23] Moreover, there is no work to compare the MFEs measured by the steady-state method with those measured by the transient method until now. We believe that a comprehensive study of the MFEs by using the steady-state method and transient method, simultaneously, would lead to an important insight into the mechanisms of MFEs.

In this work, we measured the MELs of the same devices by using both the steady-state and transient methods. The OLEDs with the structure of indium tin oxide (ITO)/*N,N*-di-1-naphthyl-*N,N*-diphenylbenzidine (NPB, 40nm)/tri-8-hydroxyquinoline-aluminum ($Alq_3$, 50 nm)/LiF (0.8 nm)/Al were fabricated by the multiple-source organic molecular beam deposition method. NPB was the hole-tansporting layer, and $Alq_3$ acted as both the electron-transporting and light-emitting layer. Immediately after being fabricated, the devices were placed on a Teflon stage between the poles of an electromagnet for the MELs measurement. In steady-state measurement, a Keithley 2612 source-measure unit was used to provide constant voltage from one channel. The light output of the devices was collected by a lens coupled with the optic fiber (2 m) connected to a Hamamatsu photomultiplier (H5783P–01, time resolution: 0.78 ns). The photomultiplier was placed far away from the electromagnet, and was connected to another channel of keithley 2612 to record the EL signals. The setup of the transient test was the same as our previous work.[23] All the measurements were carried out at room temperature under ambient condition. It should be mentioned that the performance drift of the devices is tiny during the transient measurement, while the drift could not be ignored during the steady-state test.

Fig. 1 (a) depicts the MEL measurement in the steady-state experiment. The devices were driven by a constant voltage as the magnetic field increases from 5 mT to 145 mT at a step of 5 mT. As can be seen, the EL intensity has a little increase with the driving time. In order to minimize the error from the EL drift, the zero-magnetic field (lasting for 5 s) was inserted into each step of B-field (lasting for 5 s). The MELs were calculated by: $MEL=\Delta EL/EL=[EL_{(B,N)}-EL_{(0,N)}]/EL_{(0,N)}$, where $EL_{(B,N)}$ is the EL intensity at the last point of B-field duration in the Nth step, $EL_{(0,N)}$ is the EL intensity at the first point of zero-field duration in the Nth step (see the insert of figure 1 (a)). Fig. 1 (b) shows the transient measurement in which the devices were driven by a pulse voltage with the magnetic field being on and off, respectively. In the transient test, we averaged the EL signals at the flat region (see the insert of figure 1(b)) to acquire EL(B) and EL(0), then obtained the MELs.

Figure 2 shows the MELs of the devices as a function of the external magnetic field. The devices were driven by the constant voltages ranging from 6 V to 14 V. As can be seen, the MELs increase rapidly at low magnetic field (< 30 mT), while reach saturation at high magnetic field (> 50 mT). The MELs at the magnetic field of 140 mT decease from 9.1 % to 1.9 % as the driving voltage increases from 6 V to 14 V. The curves of MELs as a function of the magnetic field can be fitted (solid line) by an empirical expression, $B^2/(B + B_0)^2$, which was suggested by Mermer *et al.*[2] The fitting parameter $B_0$ is related to the strength of the hyperfine field and is about 5 mT in our experiments, which is in good agreement with the literature data.[2,24]

Figure 3 shows the MELs of the devices as a function of the external magnetic field driven by the pulse voltages. At low driving voltages (< 7 V), the MELs increase rapidly (< 40 mT) then reach saturation (> 50mT) as the magnetic field increases, which is similar with that shown in figure 2. However, at high driving voltages (> 7 V), the results show the great difference from those obtained by the steady-state method. The MELs increase and reach maximum (at ~ 40 mT) with the augment of the magnetic field, but decline as the magnetic field increases further. Negative MELs can be seen when both the magnetic field and driving voltage are high enough.

The decrease of the MELs at high magnetic field has been reported,[1,25,26] and several mechanisms were suggested to explain it. Kalinowski *et al.* attributed the decrease of the MELs at high magnetic field to the level crossing.[1] It was said that in absence of the external magnetic field, the three substates of the triplet e-h pairs are degenerated and mix with the singlet e-h pairs due to the hyperfine interaction, singlet could convert to anyone of three substates of the triplet. When the external magnetic field is present and larger than the hyperfine field, the degeneracy of the triplet substates is removed, leading to the reduced mixing, thus the singlet population increase, resulting in the positive MEL. Generally, the e-h exchange interaction causes the energy levels of the triplet e-h pairs are below the energy level of the singlet e-h pairs. When the external magnetic field increases toward the level crossing field *B=2Jgβ* (here, *2J* is the strength of exchange interaction, *g* the *lande* factor of electron and hole, *β* the *bohr* magneton.) of $T_{+1}$ and S, a increase of the conversion from S to $T_{+1}$ can be expected, leading to the decrease of the MEL. Nevertheless, this could not explain the decrease of the MELs in our experiments. Firstly, it could not explain why the decrease of the effects was only observed in the transient experiment; secondly, from the level crossing mechanism one could not expect a negative MEL which was indeed observed in our experiment. Yan *et al.* obtained the negative MEL when they introduced a insulating layer between the anode and the hole transport layer.[25] They proposed that the unbalance injection caused by the insulating layer could increase the triplet-polaron interaction (TPI) and generate more secondary carriers. Due to that the secondary carriers trend to form singlet and the magnetic field can perturb their recombining process, leading to the decrease of the singlet population, the negative MEL can be observed. However, the injection conditions of our devices under constant voltage and pulse voltage are identical, if the negative MELs can be observed in the transient experiment, same effects should be observed in the steady-state experiment. So we think it can not be the mechanism of our negative MELs. *Δg* mechanism has been suggested to explain the negative MEL,[26] but it can not explain our results: (i) the *g* factors of the electrons and holes are almost identical in the $Alq_3$ thin film, leading to a tiny contribution of the *Δg* mechanism to the MEL under the magnetic field of about 150 mT;[26] (ii) if the Δg mechanism dominate the MELs of our device, negative MEL could be seen both in transient experiment and steady-state experiment.

A possible mechanism of our negative MELs would be the triplet-triplet annihilation (TTA) model. The MFEs on the TTA process was investigated extensively in 1970s,[27,28] and it has been introduced to explain the negative MEL in OLEDs recently by several groups.[9-11,17] There are two channels of the TTA: (a) $T + T = T^* + S^0$ (triplet channel); (b) $T + T = S^1 + S^0$ (singlet channel). The singlet channel of TTA could be suppressed by a high external magnetic field,[28] leading to a decrease of the singlet population, resulting in the negative MEL. However, one would ask why the negative MEL could not be observed in our steady-state experiments. It has been proposed that a local morphological change in the thin film and more traps could be induced by "electric conditioning" (device driven by a relatively large current for a few minutes).[13,29] We assume that the constant voltage cause an increase of the traps or defects in our device, thus the trapped carriers increase, and more triplets are annihilated by the trapped carriers or defects, leading to the decrease of the triplet density. Due to that TTA is quadratic proportional to the triplet density, the TTA can be ignored in the steady-state measurement at room temperature.[17] Thus the decrease of the MELs in our steady-state experiment vanishes. In addition, the increase of the traps can lead to the increase of the lifetime of the e-h pairs ("action centers" in ref. 29), resulting in that more e-h pairs are involved in the intersystem conversion (ISC) caused by the hyperfine interaction. This could explain the larger MELs tested by steady-state method.

In summary, we measured the MELs in the Alq$_3$ based devices through both the transient method and steady-state method. The MELs acquired from the two methods show the great different behaviors. In the steady-state experiment, the MELs were similar with the literature data. However, in the transient test, the MELs firstly increase with the external magnetic field, but decrease as the magnetic field increases further for the devices driven by a high voltage. The results indicate that the driving conditions have large impact on the MFEs in OLEDs, and the constant-voltage operation may cause additional traps in the devices. The TTA model was introduced to explain the negative MELs. Further studies of the impact of the driving conditions on the MFEs are in progress.

We are grateful for financial support from National Natural Science Foundation of China (grant numbers 60878013).

**Figure captions**

**Fig. 1** The MELs measurements of our device: (a) the EL response of the devices driven by a constant voltage (7 V) with the increasing the magnetic field; (b) the EL response of the devices driven by a pulse voltage (7 V with the repetition of 1 KHz and pulse width of 10μs) with the magnetic field (150 mT) on and off. The insert shows the enlarged EL signals at the flat region.

**Fig. 2** The MELs of the devices as a function of the external magnetic field driven by the constant voltages ranging from 6 V to 14 V (dotted), and the fitting results (solid line) by using the empirical expression of $B^2/(B + B_0)^2$

**Fig. 3** The MELs of the devices as a function of the external magnetic field driven by the pulse voltages ranging from 6 V to 20 V (repetition:1 KHz, pulse width: 10 μs).

**Fig. 1**

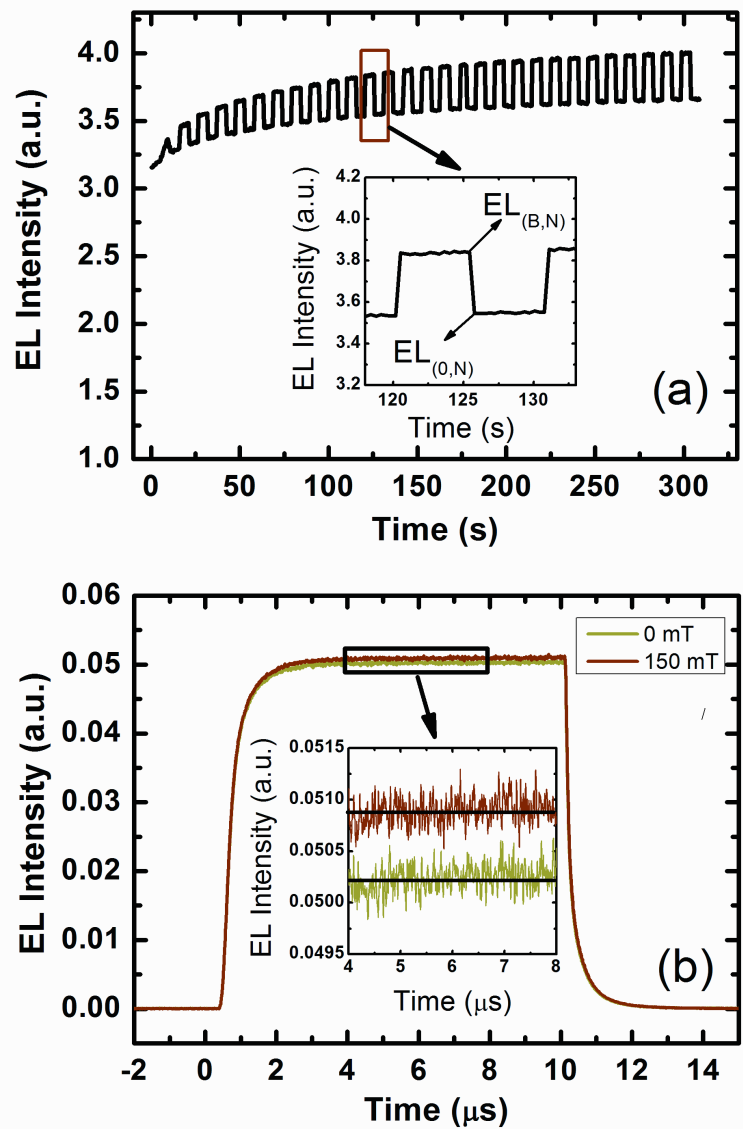

**Fig. 2**

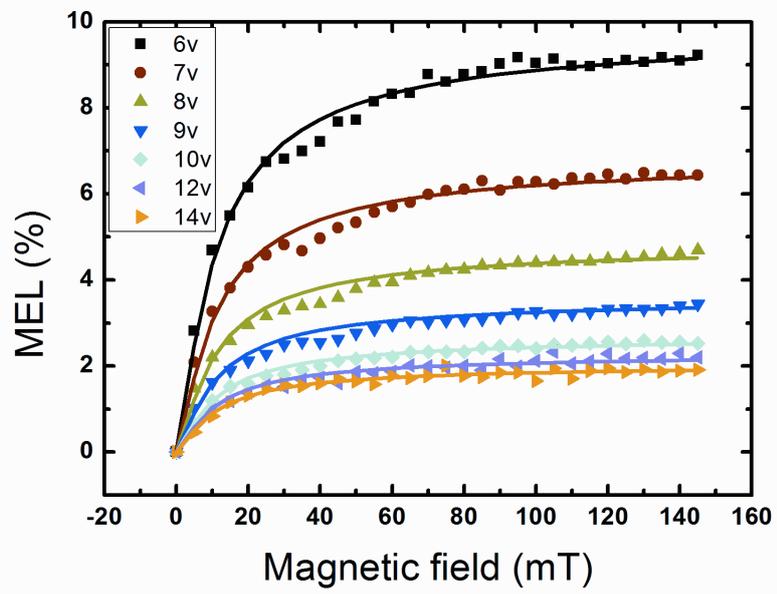

**Fig. 3**

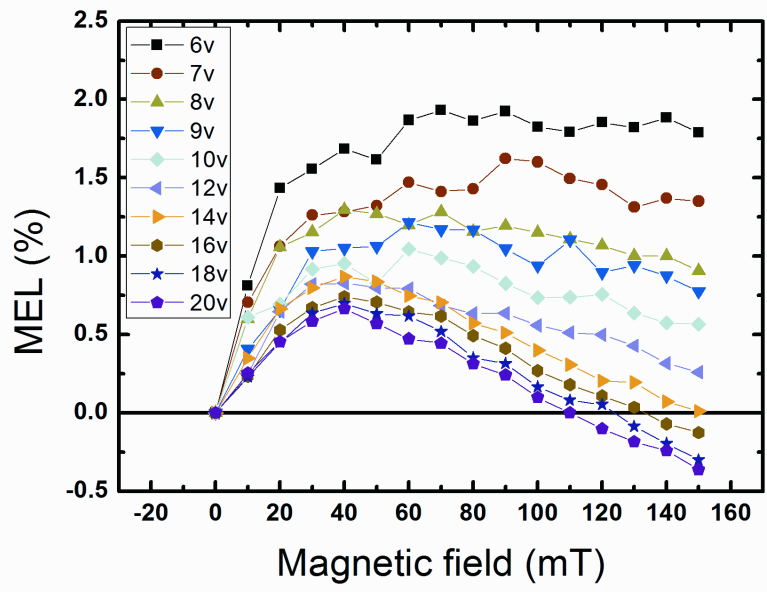